\documentclass[eqsecnum,showpacs,prc,aps,floatfix,superscriptaddress,showkeys,groupedaddress,citeautoscript]{revtex4}
\usepackage[dvips]{graphicx} 
\usepackage{graphics}
\usepackage{graphicx}
\usepackage{epsfig}
\usepackage{slashed}
\usepackage[english]{babel}
\usepackage{amsmath}
\usepackage{amssymb}
\usepackage{amssymb,amsmath,amsthm,amssymb,multirow,printlen,layouts}
\usepackage{color}
\definecolor{lightgray}{gray}{0.5}
\usepackage{xcolor}

\newcommand\nn{\nonumber}
\newcommand\ba{\begin{eqnarray}}
\newcommand\ea{\end{eqnarray}}

\begin{document}
\title{Analytical bound-state solutions of the Klein-Fock-Gordon equation for the sum of Hulth\'en and  Yukawa potential within SUSY quantum mechanics}

\author{A. I. Ahmadov}
\email{ahmadovazar@yahoo.com}
\affiliation{Department of Theoretical Physics, Baku State University, 1148 Baku, Azerbaijan}
\affiliation{Institute for Physical Problems, Baku State University, 1148, Baku, Azerbaijan}

\author {S. M. Aslanova}
\affiliation{Department of Theoretical Physics, Baku State University, 1148 Baku, Azerbaijan}

\author {M. Sh. Orujova}
\affiliation{Azerbaijan State University of Economics, Istiglaliyyat st. 22, 1001 Baku, Azerbaijan}

\author{S. V. Badalov}
\email{sabuhi.badalov@uni-paderborn.de}
\affiliation{Lehrstuhl f\"ur Theoretische Materialphysik, Universit\"at Paderborn, 33095 Paderborn, Germany}

\date{\today}

\begin{abstract}
\begin{center}\large
{Abstract}
\end{center}
The relativistic wave equations determine the dynamics of quantum fields in the context of quantum field theory. One of the conventional tools for dealing with the relativistic bound-state problem is the Klein-Fock-Gordon equation. In this work, using a developed scheme, we present how to surmount the centrifugal part and solve the modified Klein-Fock-Gordon equation for the linear combination of Hulth\'en and Yukawa potentials. In particular, we show that the relativistic energy eigenvalues and corresponding radial wave functions are obtained from supersymmetric quantum mechanics by applying the shape invariance concept. Here, both scalar potential conditions, which are whether equal and non-equal to vector potential, are considered in the calculation. The energy levels and corresponding normalized eigenfunctions are represented as a recursion relation regarding the Jacobi polynomials for arbitrary $l$ states. Beyond that, a closed-form of the normalization constant of the wave functions is found. Furthermore, we state that the energy eigenvalues are quite sensitive with potential parameters for the quantum states. The non-relativistic and relativistic results obtained within SUSY QM overlap entirely with the results obtained by ordinary quantum mechanics, and it displays that the mathematical implementation of SUSY quantum mechanics is quite perfect.

\vspace*{0.5cm}

\noindent
\end{abstract}

\maketitle

\section{INTRODUCTION}
The exactly solvable problems for quantum systems have long been a subject of intense study in many branches of quantum physics. The main aim of an analytical solution of wave equations for this attention is that the wave function contains all the requisite information for the full description of a quantum system.\cite{Greiner,Bagrov,Gara,Boivin,Iwo,Mili} In physics, especially the relativistic quantum mechanical applications to particle and nuclear physics, the relativistic wave equations predict particles' reaction at high energies.\cite{Greiner,Bagrov,Herman} The analytical solution of the Klein-Fock-Gordon (KFG) equation with physical potentials plays a central role in relativistic quantum mechanics since this wave equation perfectly defines the spinless pseudo scalar pions and Higgs boson.

In principle, numerous methods were developed, and they are still successfully implemented in solving the non-relativistic and relativistic wave equations with some familiar potentials. The Nikiforov-Uvarov method,\cite{Nikiforov} factorization method,\cite{Dong1} Laplace transform approach,\cite{Arda} and the path integral method,\cite{Cai} and shifted 1/N expansion approach\cite{Tang,Roy} for solving radial and azimuthal parts of the wave equations exactly or quasi-exactly in $l\neq0$ for various potentials. Additionally, there are numerous interesting research works about the KFG equation with physical potentials by using different methods in the literature.\cite{Hamzavi,Hartmann,Lutfuoglu1,Lutfuoglu2,Badalov0,Parmar,Gao,Zhang,Znojil,Adame,Chen,Talukar,Chetouani} Among them, as an example, in Ref.\cite{Znojil}, the s-wave KFG equation with the vector Hulth\'en type potential was treated by the standard method. As reported by Talukdar $et$ $al$, the scattering state solutions of the s-wave KFG equation with the vector and scalar Hulth\'en potentials were obtained for the irregular and regular boundary conditions.\cite{Talukar} Besides, the supersymmetry method (SUSY) was also proposed for solving the wave equations analytically.\cite{Cooper1,Cooper2,Morales,Dong_el} Nonetheless, Okon $et$ $al.$ reported analytical solutions of Schr\"{o}dinger equation for the Hulth\'en-Yukawa plus inversely quadratic potential.\cite{Okon} In Ref.\cite{Mehmet,Sever,Yuan,Qiang,Dong2,Dong3,Saad,Boztosun}, the scalar potential, which is non-equal and equal to the vector potential, was supposed to get the bound states of the KFG equation for some typical potential from the ordinary quantum mechanics. Furthermore, KFG equation with the Ring-Shaped potential was investigated by Dong $et$ $al$.\cite{Dong2} If the condition where the interaction potential is insufficient to form antiparticle-particle pairs is considered, the KFG and Dirac equations can be utilized for the investigation of zero- and 1/2-spin particles, respectively.

When a particle is in a strong field, the relativistic wave equations should be considered in the quantum system. In any case, it can be corrected quickly for non-relativistic quantum mechanics. The Hulth\'en potential is one of the essential short-range potentials in physics, extensively using to describe the continuum and bound states of the interaction systems. It has been applied to several research areas such as nuclear and particle, atomic, chemical, and condensed matter physics, so analyzing relativistic effects for a particle under this potential could become significant, especially for strong coupling. The Hulth\'en potential is defined as
\ba V_{H}(r)=-\frac{Ze^{2}}{a}\cdot\frac{e^{-r/a}}{1-e^{-r/a}}\label{a1} \ea
where $Z$ and $a$ are the atomic number and the screening parameter, respectively. They determine the range for the Hulth\'en potential.\cite{Hulten1} The Yukawa potential was proposed in 1935 as an operative potential to describe the strong interactions between nucleons.\cite{Yukawa} It takes the following form
\ba V_{Y}(r)=-\frac{Ae^{-kr}}{r},\label{a2} \ea
where $A$ describing the strength of the interaction and $1/k$ its range. Unfortunately, for an arbitrary $l$-states ($l\neq 0$), the KFG equation cannot get an exact solution with these potentials due to the centrifugal term of potentials. The numerous research works reveal the SUSY QM method's power and simplicity in solving wave equations of the central and non-central potentials for arbitrary $l$ states.\cite{Badalov4,Badalov5,Ahmadov3,Ahmadov4,Ahmadov5,Ahmadov6,Ahmadov7,Ahmadov8}

In principle, the radial function nature at the origin was investigated particularly for singular potentials by Khelashvili $et$ $al.$\cite{Khelashvili1,Khelashvili2}. While the Laplace operator is portrayed in spherical coordinates, the radial wave equation's exact derivation demonstrates the perspective of a delta function term. Thus, the delta function term of the Laplace operator yields an essential contribution to the energy level. Although the various research attempts have provided satisfactory bound state energies using Hulth\'en and Yukawa potentials separately,\cite{Gara,Matthys,Ulah,Varshni1,Varshini2,Calvin,Patil} we first considered these potentials under the linear combination form.\cite{Ahmadov9} It is also worth mentioning that this potential can be use in nuclear physics to investigate the interaction between the deformed pair of the nucleus and spin-orbit coupling for the particle motion in the potential fields. Another fascinating perspective of this potential can be used as a mathematical model in the description of vibrations on the hadronic system's side, and it can constitute a convenient model for other physical situations. The investigation of the relativistic bound states in the arbitrary $l$-wave KFG equation with the linear combination of Hulth\'en and Yukawa potentials is quite interesting, and it can provide the deeper and accurate appreciations of the physical properties of the wave functions and energies in the continuum and bound states of the interacting systems. Inspired by all developments and works, in this paper, we present the solution of the relativistic radial KFG equation for the linear combination of Hulth\'en and Yukawa potentials, defined as
\ba V(r)=-\frac{V_{0}e^{-2\delta{r}}}{1-e^{-2\delta{r}}}-\frac{Ae^{-\delta{r}}}{r},~~\label{a3} \ea
where $V_{0}=2\delta Ze^{2}$, and $\delta$ is the screening parameter.

To study the system, we use an improved scheme to overcome the centrifugal term and the SUSY quantum mechanics\cite{Gendenshtein1,Gendenshtein2}. Despite our previous research effort on this potential,\cite{Ahmadov9} the investigation of this potential still needs to be clarified in detail. Accordingly, the main goal is to solve the KFG equation for the linear combination of Hulth\'en and Yukawa potentials by considering two cases, i.e., the scalar potential which is equal and unequal to vector potential by using SUSY QM. Thereby, the energy eigenvalues and corresponding radial wave functions are found for any $l$ orbital angular momentum case. Then, we compare the obtained results with the results obtained by the NU method in ordinary quantum mechanics to present the legitimacy and feasibility of this SUSY QM method. The remainder of the paper is structured as follows. In Section \ref{br}, we introduce the analytical solution of the radial KFG equation for the linear combination of Hulth\'en and Yukawa potentials from SUSY quantum mechanics. Next, the analysis of the results is presented in Section \ref{nr}. Finally, Section \ref{cr} contains the conclusions.

\section{BOUND STATE SOLUTION OF THE RADIAL KLEIN-FOCK-GORDON EQUATION}\label{br}
\subsection{\bf Implementation SUSY Quantum Mechanics}\label{sr}
Two different types of potential can be introduced into KFG equation, which contains two objects: i) the four-vector linear momentum operator and ii) the scalar rest mass. Hence, the first one is a vector potential $V$, which introduce via minimal coupling, and the second one is a scalar potential $S$, which introduce via scalar coupling\cite{Greiner}. At this moment, they allow one to introduce two types of potential coupling: the vector potential $V$ and the space-time scalar potential $S$. The natural units ($\hbar=c=1$) are set throughout this study. In the spherical coordinates systems, the KFG equation with vector potential $V(r,\theta)$ and scalar potential $S(r,\theta)$ has the form
\ba [-\nabla^{2}+(M+S(r))^{2}]\psi(r,\theta,\phi)=[E-V(r)]^{2}\psi(r,\theta,\phi),~~~~~~~~~~\label{a4} \ea
where $E$ is the relativistic energy and $M$ denotes the rest mass of the system's scalar particle. For the separation of the angular and radial parts of the wave function, in the stationary KFG equation with the linear combination of Hulth\'en and Yukawa potentials, the wave function should be utilize the following wave function
\ba \psi(r,\theta,\phi)=\frac{\chi(r)}{r}\Theta(\theta)e^{im\phi},~~~ m=0,\pm 1,\pm 2,\pm 3 ...\label{a5} \ea
and substituting this into Eq.\eqref{a4}, the radial KFG equation is defined in the following form
\ba \chi^{''}(r)+[(E^{2}-M^{2})-2(M\cdot{S(r)}+E\cdot{V(r))}+(V^{2}(r)-S^{2}(r))-\frac{l(l+1)}{r^{2}}]\chi(r)=0.\label{a6} \ea
As it is known that the KFG equation with this potential can be solved exactly using a suitable approximation scheme to surmount the centrifugal term. To solve Eq.\eqref{a6} for $l\neq0,$ we ought to approximate the centrifugal term of the Yukawa potential in this system. As a result of this, while $\delta{r} << 1$, the improved approximation scheme,\cite{Wen1,Wei,Dong6,Jia1,Greene} must be used as
\begin{gather}
\begin{aligned}
& \frac{1}{r}\approx\frac{2\delta e^{-\delta{r}}}{1-e^{-2\delta{r}}},&\\
& \frac {1}{r^{2}}\approx\frac{4\delta^{2}e^{-2\delta{r}}}{(1-e^{-2\delta{r}})^{2}}.&
\end{aligned}\label{a7}
\end{gather}
Next, the vector and scalar potential forms for the general Hulth\'en and Yukawa potentials can be considered in the following forms
\begin{gather}
\begin{aligned}
& V_{H}(r)= -\frac{V_{0}e^{-2\delta{r}}}{1-e^{-2\delta{r}}},~~~~~S_{H}(r)= -\frac{S_{0}e^{-2\delta{r}}}{1-e^{-2\delta{r}}}&\\
& V_{Y}(r)=-\frac{V_{0}'e^{-2\delta{r}}}{1-e^{-2\delta{r}}},~~~~~S_{Y}(r)=-\frac{S_{0}'e^{-2\delta{r}}}{1-e^{-2\delta{r}}}.&
\end{aligned}\label{a8}
\end{gather}
Then, Eq.\eqref{a6} becomes as
\begin{widetext}
\begin{equation}\label{a11}
\chi^{''}(r)+\biggl[(E^{2}-M^{2})+2\left(\frac{M(S_{0}+S_{0}')+E(V_{0}+V_{0}')}{1-e^{-2\delta r}}\right)e^{-2\delta r}+\left(\frac{(V_{0}+V_{0}')^2-(S_{0}+S_{0}')^{2}}{(1-e^{-2\delta r})^2}\right)e^{-4\delta r}-\frac{4l(l+1)\delta^{2}e^{-2\delta r}}{(1-e^{-2\delta r})^2}\biggr]\chi(r)=0.
\end{equation}
\end{widetext}
Thereby, the effective potential of the Hulth\'en and Yukawa potentials linear combination has the following form
\begin{eqnarray}\label{a12}
V_{\rm eff}(r)=-\frac{4\delta^{2}(\alpha^{2}+\beta^{2})e^{-2\delta{r}}}{1-e^{-2\delta{r}}}-\frac{4\delta^{2}(\gamma^{2}-\rho^{2})e^{-4\delta{r}}}{(1-e^{-2\delta{r}})^2}+\frac{4l(l+1)\delta^{2}e^{-2\delta{r}}}{(1-e^{-2\delta r})^2},~~~~~~
\end{eqnarray}
where
\begin{gather}
\begin{aligned}
& \varepsilon=\frac{\sqrt{M^{2}-E^{2}}}{2\delta}>0,~~~\alpha=\frac{\sqrt{2EV_0+2MS_0}}{2\delta}>0,&\\
&\beta=\frac{\sqrt{2EV_0^{'}+2MS_0^{'}}}{2\delta}>0,~~~\gamma =\frac{{V_0+V_0'}}{2\delta}>0,&\\
&\rho=\frac{S_0+S'_{0}}{2\delta}>0.&
\end{aligned}\label{a14}
\end{gather}
For investigation in detail, the non-relativistic limit of the formula must be studied for the energy level. When $V(r)=S(r)$, the Eq.\eqref{a4} reduces to a Schr\"{o}dinger equation for the potential $2V(r)$. Based on supersymmetric quantum mechanics, the eigenfunction of ground state $\chi_{0}(r)$ in Eq.\eqref{a6} should be in the following form
\ba \chi_{0}(r)={N}exp{(-\int W(r)dr)},\label{a15} \ea
where $N$ and $W(r)$ are normalised constant and superpotential, respectively. The connection between the supersymmetric partner potentials $V_{-}(r)$ and $V_{+}(r)$ of the superpotential $W(r)$ is as follows\cite{Cooper1,Cooper2}
\begin{gather}
\begin{aligned}
&V_{-}(r)=W^{2}(r)-W'(r),&\\
&V_{+}(r)=W^{2}(r)+W'(r).&
\end{aligned}\label{a16}
\end{gather}
The particular solution of the Riccati equation Eq.\eqref{a16} must be in the following form
\ba W(r)=-(F+\frac{Ge^{-2\delta r}}{1-e^{-2\delta r}}),\label{a56} \ea
where $G$ and $F$ are unknown constants. Having inserted Eq.\eqref{a56} into Eq.\eqref{a16} and taking into account that $V_{-}(r)=V_{\rm eff}(r)-(E^{2}-M^{2})$, we obtain
\begin{widetext}
\begin{eqnarray}\label{a18}
F^{2}+\frac{2FGe^{-2\delta{r}}}{1-e^{-2\delta{r}}}+\frac{G^{2}e^{-4\delta{r}}}{(1-e^{-2\delta{r}})^{2}}-\frac{2\delta{G}e^{-2\delta r}}{1-e^{-2\delta{r}}}-\frac{2\delta{G}e^{-4\delta{r}}}{(1-e^{-2\delta{r}})^{2}}=\nn~~~~~~~~~~~\\
=-\frac{4\delta^{2}(\alpha^{2}+\beta^{2})e^{-2\delta{r}}}{1-e^{-2\delta{r}}}-\frac{4\delta^{2}(\gamma^{2}-\rho^{2})e^{-4\delta{r}}}{(1-e^{-2\delta{r}})^{2}}+\frac{4l(l+1)\delta^{2}e^{-2\delta{r}}}{(1-e^{-2\delta r})^2}-(E^{2}-M^{2})
\end{eqnarray}
\end{widetext}
After small simplification, it can be rewritten as
\begin{widetext}
\begin{eqnarray}\label{a19}
F^{2}+\frac{(2FG-2\delta{G})e^{-2\delta{r}}}{1-e^{-2\delta{r}}}+\frac{(G^{2}-2\delta G) e^{-4\delta{r}}}{(1-e^{-2\delta r})^{2}}=\nn~~~~~~~~~~~~~~~~~~~~~~~~~~~~~~\\
=4\delta^{2}\varepsilon^{2}-\frac{4\delta^{2}(\alpha^{2}+\beta^{2})e^{-2\delta{r}}}{1-e^{-2\delta{r}}}-\frac{4\delta^{2}(\gamma^{2}-\rho^{2})e^{-4\delta{r}}}{(1-e^{-2\delta{r}})^{2}}+{4l(l+1)\delta^{2}}[\frac {e^{-2\delta{r}}}{1-e^{-2\delta{r}}}+\frac{e^{-4\delta{r}}}{(1-e^{-2\delta r})^{2}}]
\end{eqnarray}
\end{widetext}
From comparison of compatible quantities in the left and right sides of the equation Eq.\eqref{a19}, we find the following relations for $G$ and $F$ constants
\ba F^{2}=4\delta^{2}\varepsilon^{2},\label{a57} \ea
\ba 2FG-2\delta G=4\delta^{2}l(l+1)-4\delta^{2}(\alpha^{2}+\beta^{2}),\label{a58} \ea
\ba G^{2}-2\delta G=-4\delta^{2}(\gamma^{2}-\rho^{2})+4\delta^{2}l(l+1).\label{a59} \ea
Considering extremity conditions for wave functions, we obtain $G>0$ and $F<0$. Solving Eq.\eqref{a59} yields
\ba G=\delta\pm 2\delta\sqrt{(l+\frac{1}{2})^{2}-\gamma^{2}+\rho^{2}},\label{a60} \ea
and considering $G>0$ from Eqs.\eqref{a58} and \eqref{a59}, we find that
\ba F=\frac{G}{2}-\frac{2\delta^{2}(\alpha^{2}+\beta^{2}-\gamma^{2}+\rho^{2})}{G}.\label{a62} \ea
From Eq.\eqref{a57} and Eq.\eqref{a62}, we find that
\ba \varepsilon^{2}=\frac{1}{4\delta^{2}}[\frac{\delta+2\delta\sqrt{(l+\frac{1}{2})^{2}-\gamma^{2}+\rho^{2}}}{2}-\frac{2\delta(\alpha^{2}+\beta^{2}-\gamma^{2}+\rho^{2})}{1+2\sqrt{(l+\frac{1}{2})^{2}-\gamma^{2}+\rho^{2}}}]^{2}.\label{a63} \ea
After inserting the \eqref{a63} into the \eqref{a14} for the definitions the energy eigenvalue of ground state for general case $V(r)\neq S(r)$, we obtain the following energy level equation
\ba M^{2}-E_{0}^{2}=[\frac{\delta+2\delta\sqrt{(l+\frac{1}{2})^2-\gamma^{2}+\rho^{2}}}{2}-\frac{2\delta(\alpha^{2}+\beta^{2}-\gamma^{2}+\rho^{2})}{1+2\sqrt{(l+\frac{1}{2})^2-\gamma^{2}+\rho^{2}}}]^2.\label{a64} \ea
When $r\rightarrow{\infty}$, the chosen superpotential $W(r)\rightarrow$-$F$. Inserting the Eq.\eqref{a56} into Eq.\eqref{a16}, the supersymmetric partner potentials $V_{-}(r)$ and
$V_{+}(r)$ can be found in the following forms
\begin{gather}
\begin{aligned}\label{a65}
&V_{-}(r)=F^{2}+\frac{(2FG-2\delta{G})e^{-2\delta{r}}}{1-e^{-2\delta{r}}}+\frac{(G^{2}-2\delta{G})e^{-4\delta{r}}}{(1-e^{-2\delta{r}})^2},&\\
&V_{+}(r)=F^{2}+\frac{(2FG+2\delta{G})e^{-2\delta{r}}}{1-e^{-2\delta{r}}}+\frac{(G^{2}+2\delta{G})e^{-4\delta{r}}}{(1-e^{-2\delta{r}})^2}.&
\end{aligned}
\end{gather}
By using the superpotential $W(r)$ from Eq.\eqref{a56}, we can find $\chi_{0}(r)$ radial eigenfunction of ground state in the following form
\begin{equation}\label{a67}
\chi_{0}(r)=Ne^{Fr}(1-e^{-2\delta{r}})^{\frac{G}{2\delta}}
\end{equation}
where $r\rightarrow{0}$; $\chi_{0}(r)\rightarrow{0}$, $G>0$, and $r\rightarrow \infty$; $\chi_{0}(r)\rightarrow0, F<0$. Two partner potentials $V_{-}(r)$ and $V_{+}(r)$ which differ from each other with additive constants and have the same functional form are called the invariant potentials.\cite{Gendenshtein1, Gendenshtein2} Hence, for the partner potentials $V_{-}(r)$ and $V_{+}(r)$ given with Eq.\eqref{a16} and Eq.\eqref{a56}, the invariant forms are defined as
\begin{widetext}
\begin{equation}\label{a68}
R(G_{1})=V_{+}(G,r)-V_{-}(G_{1},r)=F^{2}-F_{1}^{2}=[\frac{G}{2}-\frac{2\delta^{2}(\alpha^{2}+\beta^{2}-\gamma^{2}+\rho^{2})}{G}]^{2}-[\frac{G+2\delta}{2}-\frac{2\delta^{2}(\alpha^{2}+\beta^{2}-\gamma^{2}+\rho^{2})}{G+2\delta}]^{2},
\end{equation}
\end{widetext}
\begin{widetext}
\begin{eqnarray}\label{a69}
R(G_{i})=V_{+}[G+(i-1)2\delta ,r]-V_{-}[G+i2\delta ,r]=\nn~~~~~~~~~~~~~~~~~~~~~~~~~~~~~~\\
=(\frac{G+(i-1)\cdot2\delta}{2}-\frac{2(\alpha^{2}+\beta^{2}-\gamma^{2}+\rho^{2})\delta^2}{G+(i-1)\cdot2\delta})^{2}-(\frac{G+i\cdot2\delta}{2}-\frac{2(\alpha^{2}+\beta^{2}-\gamma^{2}+\rho^{2})\delta^{2}}{G+i\cdot2\delta})^{2}.
\end{eqnarray}
\end{widetext}
where the reminder $R(G_i)$ is independent of $r$. If we keep going this procedure and make the following substitution $\,G_{n_r}=G_{n_r-1}+2\delta=G+2n_r\delta$, the whole discrete level of Hamiltonian $\,H_{-}(G)$ can be written as
\ba E^{2}_{n_r}=E^{2}_0+\sum\limits_{i=1}^{n} R(G_i),\label{a71} \ea
and we obtain the following form
\ba E^{2}_{n_{r}l}&=&M^{2}-(\frac{G+2\delta{n_{r}}}{2}-\frac{2\delta^{2}(\alpha^{2}+\beta^{2}-\gamma^{2}+\rho^{2})}{G+2\delta{n_{r}}})^{2}.\label{a72}~~~~~~~~~~~~ \ea
In the following, we obtain the energy level equation in ordinary quantum mechanics
\begin{widetext}
\begin{eqnarray}\label{a73}
M^{2}-E_{n_{r},l}^{2}=[\frac{\alpha^{2}+\beta^{2}-(l+\frac{1}{2})^{2}-(n_r+\frac{1}{2})^{2}-2(n_{r}+\frac{1}{2})\sqrt{(l+\frac{1}{2})^{2}-\gamma^{2}+\rho^{2}}}{n_r+\frac{1}{2}+\sqrt{(l+\frac{1}{2})^{2}-\gamma^{2}+\rho^{2}}}\cdot\delta]^{2}.
\end{eqnarray}
\end{widetext}
As seen from Eq.\eqref{a73}, it is in a perfect agreement with the result obtained in Eq.(27) of Ref.\cite{Ahmadov8}. If we consider Eq.\eqref{a14} into Eq.\eqref{a73} and do some simple algebraic derivation, we can obtain the energy eigenvalues equation in the simplest form
\begin{widetext}
\begin{eqnarray}\label{a732}
M^{2}-E_{n_{r}l}^{2}=[{\delta(n_{r}+\frac{1}{2}+\sqrt{(l+\frac{1}{2})^{2}-\gamma^{2}+\rho^{2}})-\frac{\gamma E_{n_{r}l}+\rho{M}+\delta(\rho^{2}-\gamma^{2})}{n_{r}+\frac{1}{2}+\sqrt{(l+\frac{1}{2})^{2}-\gamma^{2}+\rho^{2}}}}]^{2}
\end{eqnarray}
\end{widetext}
with $\alpha^{2}+\beta^{2}=\frac{\gamma{E_{n_{r}l}}+\rho{M}}{\delta}$.

\begin{figure*}
\includegraphics[width=\textwidth]{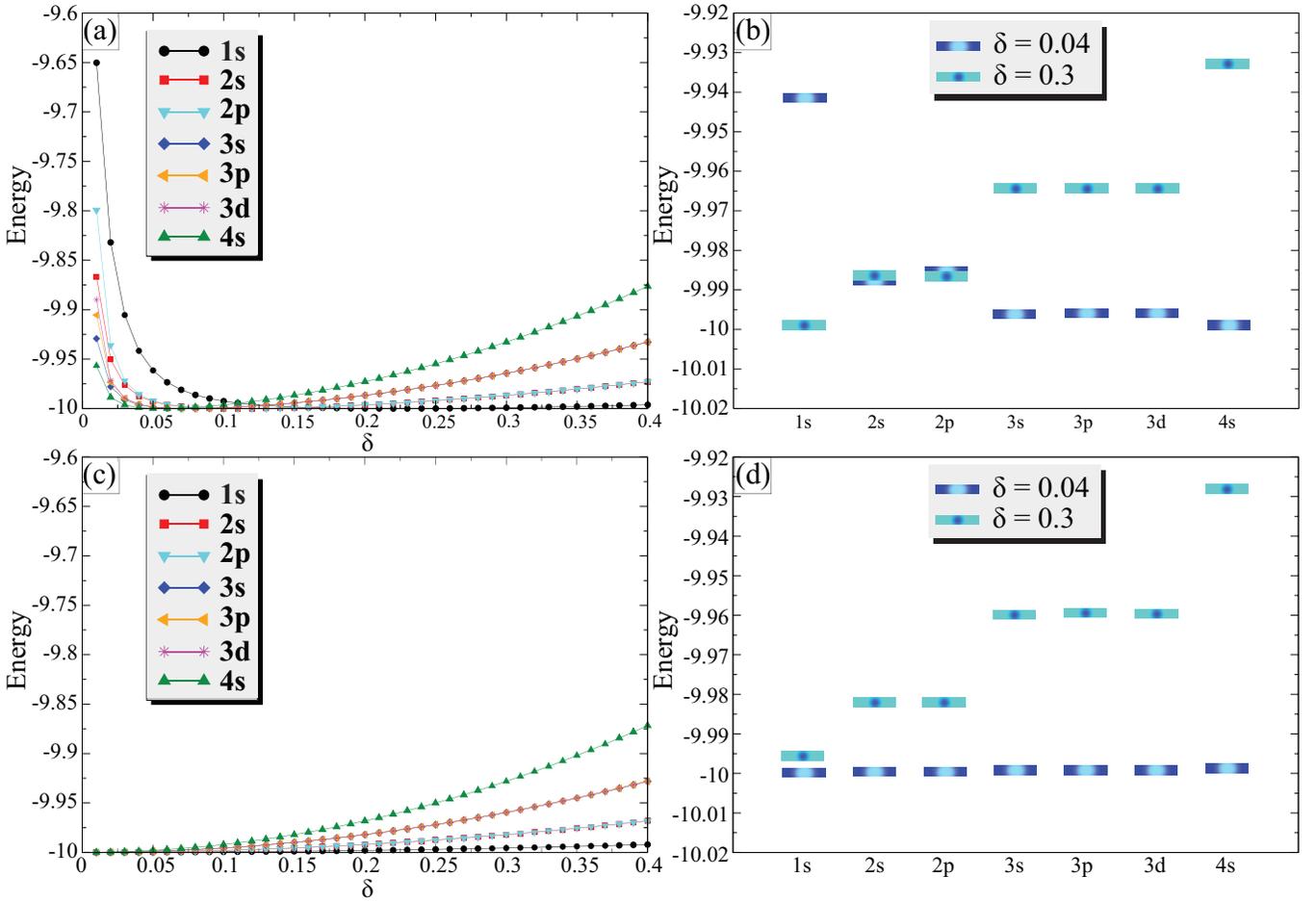}
\caption{\label{Fig1}(Color online) The variation of energy level as a function of screening parameter $\delta$ for quantum states in {(a,b)} the parameters $M$=10, $V_{0}$=0.01, $V{'}_{0}$=0.05, $S_{0}$=0.025, $S{'}_{0}$=0.035 and {(c,d)} the parameters $M$=10, $V_{0}$=0.02, $V{'}_{0}$=0.06, $S_{0}$=0.035, $S{'}_{0}$=0.045.}
\end{figure*}

Based on the SUSY QM method and knowing the ground state eigenvalues $E_{0}$ and eigenfunctions $\chi_{0}$, all energy eigenvalues $E_{{n_{r}{l}}}$ and eigenfunctions $\chi_{{n_{r}{l}}}$ can be easily obtained. Briefly, using the following equation
\ba
\chi_{n_{r}}(r,a_{0})=A^{+}(r,a_{0})\chi_{n_{r}-1}(r,a_{1}),
\label{a746} \ea
$\chi_{{n_{r}{l}}}$ can be easily obtained in terms of the  ground  state  wave functions. The superpotential $W(r)$ depends on two parameters ${a_{0}}=(F,G)$ and the first partner potential has like that parameter ${a_{1}}=(F_{1},G_{1})$. Hence, Eq.\eqref{a746} will be in the following form
\ba
\chi_{n_{r}}(r,a_{0})=(-\dfrac{d}{dr}-F-\dfrac{Ge^{-2\delta{r}}}{1-e^{-2\delta{r}}})\chi_{n_{r}-1}(r,a_{1}),
\label{a747} \ea
We define a new variable $s=e^{-2\delta r}\in[0,1]$ and factoring out the ground state wavefunction
\ba
\chi_{n_{r}}(s,a_{0})=\chi_{0}(s,a_{0})R_{n_{r}}(s,a_{0}).
\label{a748} \ea
Substituting into Eq.\eqref{a747} and using the ground state wavefunction Eq.\eqref{a67}, we get
\ba
R_{n_{r}}(s;\epsilon, K)=s(1-s)\dfrac{d}{ds}R_{n_{r}-1}(s;\epsilon, K+1)+[2\epsilon-(2{\epsilon}+2K+1)]R_{n_{r}-1}(s;\epsilon, K+1).
\label{a748} \ea
Based on comparison it with the recursion relation in Ref:\cite{Abramowitz}
\ba
P_{n_{r}}^{(\alpha,\beta)}(1-2s)=s(1-s)\dfrac{d}{ds}P_{n_{r}-1}^{({{\alpha}+1},{{\beta}+1})}(1-2s)+[{\alpha}+1-({\alpha}+{\beta}+2)s)]P_{n_{r}-1}^{({{\alpha}+1},{{\beta}+1})}(1-2s),
\label{a749} \ea
it is seen that $R_{n_{r}}(s,a_{0})$ is proportional to  the  Jacobi  polynomial $P_{n_{r}}^{(2\epsilon,2K-1)}(1-2s)$. Thus, the normalized eigenfunction for this potential is taken in the following form
\ba \chi_{{n_{r}{l}}}(s)=C_{{n_{r}{l}}}s^{\varepsilon}(1-s)^{K}P_{n_{r}}^{(2\epsilon,2K-1)}(1-2s),\label{a50_1} \ea
or
\ba \chi_{{n_{r}{l}}}(s)=C_{{n_{r}{l}}}s^{\varepsilon}(1-s)^{K}\frac{\Gamma(n_{r}+2\varepsilon+1)}{n_{r}!
\Gamma(2\varepsilon+1)}\cdot{_{2}F_{1}}\left({-n_{r},{2\varepsilon}+2K+n_{r},1+{2\varepsilon};s}\right),\label{a50} \ea
where $K=1/2+\sqrt{(l+1/2)^{2}-\gamma^{2}-\rho^{2}}$.
The normalization constant $C_{n_{r}l}$ can be found by using the normalization condition
\ba \int\limits_0^{\infty}|R(r)|^{2}r^{2}dr=\int\limits_0^{\infty}|\chi(r)|^{2}dr=\frac{1}{2\delta}\int\limits_0^1\frac{1}{s}|\chi(s)|^{2}ds=1,
\label{a51} ~~~~~~~~~\ea
by utilizing the following integral formula in Ref.\cite{Abramowitz}:

\begin{widetext}
\begin{eqnarray}\label{a52}
\int\limits_0^{1}{(1-z)^{2(\delta+1)}z^{{2\lambda}-1}}{{_{2}F_{1}(-n_{r},2(\delta+\lambda+1)+n_{r},2\lambda+1;z)}}^{2} dz=  \nn  \\ =\frac{{(n_{r}+{\delta}+1)n_{r}!\Gamma(n_{r}+{2\delta}+2)\Gamma(2\lambda)\Gamma({2\lambda}+1)}}{{(n_{r}+{\delta}+{\lambda}+1)
\Gamma(n_{r}+{2\lambda}+1)\Gamma(2({\delta}+{\lambda}+1)+n_{r})}},~~~~~~~~
\end{eqnarray}
\end{widetext}
where $\lambda>0$ and $\delta>-\frac{{3}}{2}$.
After making simple calculations, we arrive at the following expression for the normalization constant
\ba C_{{n_r}l}=\sqrt{\frac{2\delta n_{r}!(n_{r}+K+\varepsilon)\Gamma({2\varepsilon}+1)\Gamma(n_{r}+{2\varepsilon}+2K)}{(n_{r}+K)\Gamma(2\varepsilon)\Gamma(n_{r}+2K)\Gamma(n_{r}+{2\varepsilon}+1)}}.
\label{a53}~~~~~~~~~~~~ \ea

\section{\bf Results and Discussion}\label{nr}
In this section, we present the numerical evaluation for the bound state solutions of the l-wave KFG equation with the vector and scalar form of the linear combination of Hulth\'en and Yukawa potentials. To study the property of the energy levels regarding potential parameters in some quantum states, (see Figure \ref{Fig1}) we take $M$=10, $V_{0}$=0.01, $V{'}_{0}$=0.05, $S_{0}$=0.025, $S{'}_{0}$=0.035 and  $M$=10, $V_{0}$=0.02, $V{'}_{0}$=0.06, $S_{0}$=0.035, $S{'}_{0}$=0.045. The little difference ($\sim$0.01) in model potential parameters $V_{0}$, $V{'}_{0}$, $S_{0}$, $S{'}_{0}$ is quite sufficient, in order to see the energy level of quantum states displayed completely different behavior.
In the Figure \ref{Fig1} (a, b), the energy levels $E$ of quantum states first are decreasing until some of small $\delta$ values ($\sim$0.075-0.1), then the energy levels $E$ increase in the $\delta$>0.1. In the Figure \ref{Fig1} (c, d), the energy levels $E$ of quantum states have very little variation for an interval of $\delta\in[0,0.1]$, and it causes the degenerate of all quantum states, then the energy levels $E$ of quantum states continue to gradually increase with increments of $\delta$. These behaviors are better recognized in higher quantum states.

Behind that of these results, we can investigate some special cases.

i) In case $S_{0}=V_{0}$, and $S_{0}^{'}=V_{0}^{'}$, namely $\gamma=\rho$, we obtain as
\ba M^{2}-E_{n_{r}l}^{2}=[{\delta(n_{r}+l+1)-\frac{\gamma(E_{n_{r}l}+M)}{n_{r}+l+1}}]^{2},\label{a74}
\ea where $\gamma=\frac{V_{0}+V_{0}^{'}}{2\delta}$.

ii) If $V_{0}^{'}=0$, $S_{0}^{'}=0$, we obtain the energy level equation for Hulth\'en potential case.
\begin{widetext}
\begin{eqnarray}\label{a734}
M^{2}-E_{n_{r}l}^{2}=[{\delta(n_{r}+\frac{1}{2}+\sqrt{(l+\frac{1}{2})^{2}-{\gamma'}^{2}+{\rho'}^{2}})-\frac{{\gamma'}E_{n_{r}l}+{\rho'}M+\delta({\rho'}^{2}-{\gamma'}^{2})}{n_{r}+\frac{1}{2}+\sqrt{(l+\frac{1}{2})^{2}-{\gamma'}^{2}+{\rho'}^{2}}}}]^{2},
\end{eqnarray}
\end{widetext}
where ${\gamma'}=\frac{V_{0}}{2\delta}$ and ${\rho'}=\frac{S_{0}}{2\delta}$. This result is in good agreement with the expression obtained in Eq.(50) of Ref:\cite{Ahmadov9}.

iii) In case $V_{0}=0$, $S_{0}=0$, but $V_{0}^{'}\neq S_{0}^{'}$, we obtain the energy level equation for Yukawa potential, which defined as following form
\begin{widetext}
\begin{eqnarray}\label{a735}
M^{2}-E_{n_{r}l}^{2}=[{\delta(n_{r}+\frac{1}{2}+\sqrt{(l+\frac{1}{2})^{2}-{\gamma''}^{2}+{\rho''}^{2}})-\frac{{\gamma''}E_{n_{r}l}+{\rho''}M+\delta({\rho''}^{2}-{\gamma''}^{2})}{n_{r}+\frac{1}{2}+\sqrt{(l+\frac{1}{2})^{2}-{\gamma''}^{2}+{\rho''}^{2}}}}]^{2},\label{a736}
\end{eqnarray}
\end{widetext}
where ${\gamma''}=V'_{0}/2\delta$, and ${\rho''}=S'_{0}/2\delta$.
This result is in good agreement with the expression obtained in Eq.(52) of Ref:\cite{Ahmadov9}. Furthermore, this result is also the same with the expression for the constant mass case obtained in Ref:\cite{Wang}. One can easily see this by setting $q$=1 and ${\alpha}\rightarrow{\delta}$ in Eq.(39) of Ref:\cite{Wang}.

iv) Also, if $V_{0}=-S_{0}$, and $V_{0}^{'}=-S_{0}^{'}$, namely $\gamma=-\rho$.
\ba M^{2}-E_{n_{r}l}^{2}=[{\delta(n_{r}+l+1)-\frac{\gamma(E_{n_{r}l}-M)}{n_{r}+l+1}}]^{2},\label{a78}
\ea where $\gamma=\frac{V_{0}+V_{0}^{'}}{2\delta}$.

v) If $\delta\rightarrow 0$ and $S_{0}=V_{0}=2\delta{Z}e^{2}$ or $\gamma^{'}=\rho^{'}=\dfrac{V_{0}}{2\delta}=Ze^{2}$ in Eq.\eqref{a732}, the potential reduces to Coulomb potential, $V_{c}(r)=-Ze^{2}/r$, and the corresponding energy spectrum is obtained as
\ba E_{n_{r}l}=\dfrac{{(n_{r}+l+1)}^{2}-Z^{2}e^{4}}{{(n_{r}+l+1)}^{2}+Z^{2}e^{4}}M
\label{a781} \ea
and this result is the same with Eq.(51) of Ref:\cite{Wang}.

vi) If we take $l=0$ (the s-wave case), the centrifugal term in Eq.\eqref{a11} disappears because $\frac{4l(l+1)\delta^{2}e^{-2\delta{r}}}{(1-e^{-2\delta{r}})^2}=0$ and the equation turns to the s-wave KFG equation. By setting $l=0$ in Eq.\eqref{a732}, its energy spectrum equation is the following form
\begin{widetext}
\begin{eqnarray}\label{a744}
M^{2}-E_{n_{r}l}^{2}=[{\delta(n_{r}+\frac{1}{2}+\sqrt{\dfrac{1}{4}-\gamma^{2}+\rho^{2}})-\frac{\gamma E_{n_{r}l}+\rho{M}+\delta(\rho^{2}-\gamma^{2})}{n_{r}+\frac{1}{2}+\sqrt{\dfrac{1}{4}-\gamma^{2}+\rho^{2}}}}]^{2}
\end{eqnarray}
\end{widetext}

vii) If we take $S_{0}$=$V_{0}$ and $S'_{0}$=$V'_{0}$, and based on the following transformations $E_{n_{r}l}$--$M$ $\rightarrow$ $E_{n_{r}l}^{^{NR}}$,
 $E_{n_{r}l}$+$M$ $\rightarrow$ $2M$, ~~~$V_{0}$ $\rightarrow$ $\frac{V_{0}}{2}$, and $V'_{0}$ $\rightarrow$ $\frac{V'_{0}}{2}$, we obtain the energy level equation of Eq.\eqref{a732} for the non-relativistic case. Briefly, because of the following relation
\begin{gather}
\begin{aligned}
&M^{2}-E_{n_{r}l}^{2}=(M-E_{n_{r}l})(M+E_{n_{r}l})=-E_{n_{r}l}^{^{NR}}\cdot{2M},& \\
&\gamma=\rho=\frac{V_{0}+V_{0}^{'}}{4\delta}=\frac{V_{0}+2\delta{A}}{4\delta},& \\
&\gamma(E_{n_{r}l}+M)=2M\gamma=\frac{M(V_{0}+2\delta{A})}{2\delta},&
\end{aligned}\label{eq:a745}
\end{gather}
the energy level equation of Eq.\eqref{a732} for the non-relativistic case, can be written the following form
\begin{equation}
E_{n_{r}l}^{^{NR}}=-\frac{1}{2M}\left[\delta(n_{r}+l+1)+\frac{M(V_{0}+2\delta{A})}{2\delta(n_{r}+l+1)}\right]^{2},
\end{equation}
which is good agreement with the result in Eq.(28) (If B and C are considered zero as a special case) of Ref:\cite{Okon}.
Generally, it is obviously seen from Eq.\eqref{a732} the bound states show more stability in the case of the linear combination of Hulth\'en and Yukawa potentials than Hulth\'en and Yukawa potentials cases. Furthermore, the energy eigenvalues of the quantum states are considerably sensitive with depending potential parameters.

\section{\bf Conclusion}\label{cr}
To conclude, we admit that the SUSY QM method was presented to solve the KFG equation for the linear combination of Hulth\'en and Yukawa potentials. Hence the energy eigenvalues and corresponding eigenfunctions of a mentioned quantum system were analytically obtained for arbitrary $l$ angular momentum and $n_r$ radial quantum numbers. Next, a closed-form of the normalization constant of the wave functions was also found. Beyond that, it was also shown that the energy eigenvalues are considerably sensitive respecting quantum states. Finally, the results obtained within SUSY QM are in excellent agreement with the results obtained by ordinary quantum mechanics, and it confirms that the mathematical application of SUSY quantum mechanics is ideal for similar systems.

It is worth mentioning that the main results of this paper are the explicit and closed-form expressions for the energy eigenvalues and the normalized wave functions. The method presented in this study is systematic, and in many cases, one of the most definite works in this field. In particular, the linear combination of Hulth\'en and Yukawa potentials can be one of the essential exponential potentials, and it probably provides a promising avenue in many branches of physics, especially in hadronic and nuclear physics.
\\
\\

\appendix
\section{SUSYQM Method}
For $N=2$ in SUSYQM, it is possible to define two nilpotent operators, $Q$ and $Q^{\dagger}$. They satisfy the following anti-commutation relations: 
\ba
\{ Q\, ,\, Q\}=0,\, \{Q^{\dagger},Q^{\dagger} \} =0, \nn \\
\{ Q,\, Q^{\dagger} \}=H.
\ea
%
Here $H$ is the supersymmetric Hamiltonian operator and conventionally 
$Q=\left(\begin{array}{cc}{0}&{0} \\
{A^{-}}&{0} \end{array}\right)$
and 
$Q^{\dagger}
=\left(\begin{array}{cc} {0} & {A^{+}} \\ {0} & {0}
\end{array}\right)$. The $Q$ and $Q^{\dagger}$ are also known as the supercharges operators. Here $A^{-} $ is
bosonic operator and $A^{+}$ is its adjoint. In terms of these operators, the Hamiltonian $H$ can be defined as
\cite{Cooper1,Cooper2}:
\ba H=\left(\begin{array}{cc}{A^{+}A^{-}}&{0} \\ {0}&{A^{-}A^{+}} \end{array}\right)\,
=\left(\begin{array}{cc}{H_{-}}&{0}\\{0}&{H_{+}}
\end{array}\right),
\ea where the $H_{\pm}$ are named as the Hamiltonian of supersymmetric-partner. Note also that $Q$ and $Q^{\dagger}$ operators commute with $H$. If we have zero ground state energy for $H$ (i.e. $E_{0}=0$), we can always represent the Hamiltonian as a product of a linear differential operators pairs in a factorable form. Therefore, the ground state $\psi_{0}(x)$ obeys the Schr\"{o}dinger equation as follows: \ba
H\psi_{o}(x)=-\frac{\hbar^{2} }{2m} \frac{d^{2} \psi_{0}}{dx^{2}}+V(x)\psi_{0} (x)=0, \ea hence \ba
V(x)=\frac{\hbar^{2}}{2m} \frac{\psi ''_{0} (x)}{\psi _{0}(x)}. \ea
This result makes us possible to globally reconstruct the above potential from the information of its ground state wave function that contain zero nodes. Hence, factorizing of $H$ is quite easy by
using the following ansatz \cite{Cooper1,Cooper2}: \ba
H_{-}=-\frac{\hbar^{2}}{2m} \frac{d^{2}}{dx^{2}}+V(x)=A^{+}A^{-} \ea
where \ba A^{-} =\frac{\hbar}{\sqrt{2m}} \frac{d}{dx}+W(x)\,, \,
A^{+}=-\frac{\hbar}{\sqrt{2m}} \frac{d}{dx}+W(x). \ea After that, the Riccati equation for $W(x)$ can be written as \ba
V_{-}(x)=W^{2}(x)-\frac{\hbar}{\sqrt{2m}} W'(x). \ea Solving for $W(x)$ from this equation, we can express it in terms of $\psi_{0}(x)$ by \ba W(x)=-\frac{\hbar}{\sqrt{2m}}
\frac{\psi'_{0}(x)}{\psi_{0}(x)}. \ea We obtain this solution by noticing that when $A^{-}\psi _{0} (x)=0$ is satisfied, we have $H\psi_{0}=A^{+} A^{-} \psi_{0}=0 \,.$ We then introduce the
operator $H_{+} =A^{-} A^{+} $ which is written by reversing the order of the $H^{-}$ components. After a bit simplification, we find that $H_{+}$ is nothing but the Hamiltonian for new potential
$V_{+}(x)$. \ba H_{+} =-\frac{\hbar^{2}}{2m}
\frac{d^{2}}{dx^{2}}+V_{+} (x)\, \, \, ,\, \, \, \,
V_{+}(x)=W^{2}(x)+\frac{\hbar}{\sqrt{2m}} W'(x). \ea We call $V_{\pm}(x)$ as supersymmetric partner potentials. For example, when the ground state energy of $H_{1}$ is $E^{1}_{0}$ with eigenfunction
$\psi _{0}^{1}$, from Eq.(A.5) we can always write \ba
H_{1}=-\frac{\hbar^{2}}{2m} \frac{d^{2}}{dx^{2}}+V_{1}(x)=A^{+} A^{-}+E_{0}^{1}, \ea where \ba
\begin{array}{l} {A_{1}^{-}=\frac{\hbar}{\sqrt{2m}}
\frac{d}{dx}+W_{1}(x)\, ,\, \, A_{1}^{+}=-\frac{\hbar}{\sqrt{2m}}
\frac{d}{dx}+W_{1}(x),}
\\ V_{1}(x)=W_{1}^{2}(x)-\frac{\hbar}{\sqrt{2m}} W'_{1}(x)+E_{0}^{1}, W_{1}(x)=-\frac{\hbar}{\sqrt{2m}}\frac {d \ln \psi_{0}^{1}}{dx} \,.\end{array}
\ea The SUSY partner Hamiltonian is defined by
\cite{Cooper1,Cooper2} \ba H_{2} =A_{1}^{-} A_{1}^{+}
+E_{0}^{1}=-\frac{\hbar ^{2} }{2m} \frac{d^{2} }{dx^{2} } +V_{2}
(x), \ea where \ba
\begin{split}
V_{2}(x)&=W_{1}^{2}(x)+\frac{\hbar}{\sqrt{2m}}W'_{1}(x)+E_{0}^{1} \\
&= V_{1}(x)+\frac{2\hbar}{\sqrt{2m}}
W'_{1}(x)=V_{1}(x)-\frac{\hbar^2}{m} \frac{d^{2}}{dx^{2}} (\ln
\psi_{0}^{(1)}).
\end{split}
\ea Using Eq.(A.12), for $H_{1}$ and $H_{2}$, the energy eigenvalues and eigenfunctions are obtained as \ba E_{n}^{2} =E_{n+1}^{1} \, ,\,
\, \, \,
\psi_{n}^{2}=[E_{n+1}^{1}-E_{0}^{1}]^{-\frac{1}{2}}A_{^{1}}^{-}
\psi_{n+1}^{1}\, ,\, \, \, \, \psi_{n+1}^{1}=[E_{n}^{2} -E_{0}^{1}
]^{-\frac{1}{2}}A_{^{1}}^{+}\psi_{n}^{2}. \ea Here $E_{n}^{m}$
represents the energy eigenvalue, where $n$ and $m$ denote the energy level and the  $m$'th Hamiltonian $H_{m}$, respectively. Hence, it is clear that if $H_{1}$ has $p\ge 1$ bound states with
corresponding eigenvalues $E_{n}^{1}$, as well as eigenfunctions $\psi_{n}^{1}$ defined in $0<n<p$, then we can always generate a hierarchy of $(p-1)$ Hamiltonians, i.e., $H_{2}, H_{3} \, ,\, ...,\,
H_{p}$ such that the $(H_{m})$ has the same spectrum of eigenvalue as $H_{1}$, apart from the fact that the first $(m-1)$ eigenvalues of $H$ are absent in $H$\cite{Cooper1,Cooper2}: \ba H_{m}
=A_{m}^{+} A_{m}^{-} +{\rm \; E}_{{\rm m-1}}^{{\rm
1}}=-\frac{\hbar^{2}}{2m} \frac{d^{2} }{dx^{2} } +V_{m} (x), \ea
where \ba A_{m}^{-}=\frac{\hbar}{\sqrt{2m}} \frac{d}{dx}+W_{m}(x)\,
,\, \, \, \, W_{m}(x)=-\frac{\hbar}{\sqrt{2m}} \frac{d\ln
\psi_{0}^{(m)}}{dx},  (m=2\,\,3\,\, 4,\,\, \cdots \, \, p). \ea We also have
\ba \begin{array}{l} E_{n}^{(m)}=E_{n+1}^{(m-1)}=\cdots =E_{n+m-1}^{1} \,, \\ \psi _{n}^{(m)} =[E_{n+m-1}^{1}-E_{m-2}^{1}]^{-\frac{1}{2}} \cdots [E_{n+m-1}^{1}-E_{0}^{1}]^{-\frac{1}{2}}A_{m-1}^{-} \cdots A_{1}^{-} \psi^{1}_{n+m-1}, \,\,\,\, \\
V_{m}(x)=V_{1}(x)-\frac{\hbar^2}{m} \frac{d^{2}}{dx^{2}} \ln
(\psi_{0}^{(1)} \cdots \psi _{0}^{(m-1)}). \end{array} \ea such that, by knowing all the eigenfunctions and eigenvalues of $H_{1}$ we also obtain the corresponding eigenfunctions $\psi_{n}^{1}$ and energy eigenvalues $E_{n}^{1}$ of the $(p-1)$ Hamiltonians $(H_{2}\, ,\, \, H_{3} \, ,\, ...,\, H_{p})$.

\begin{center}
{\bf Conflict of Interest Statement}
\end{center}
All authors have no conflicts of interest.

\end{document}